\documentclass[twocolumn]{pasj00}
\SetRunningHead{TSUMURA et al.}{Diffuse Galactic Light and 3.3 $\mu$m PAH with AKARI IRC} 
\Received{2013 January 30}
\Accepted{2013 July 25}
\Published{}        
\newcommand{\nw}{nWm$^{-2}$sr$^{-1}$}
\newcommand{\uw}{$\mu $Wm$^{-2}$sr$^{-1}$}

\begin{document}
\title{Low-Resolution Spectrum of the Diffuse Galactic Light and 3.3 $\mu$m PAH emission with AKARI InfraRed Camera} 
\author{Kohji \textsc{Tsumura}\altaffilmark{1}, Toshio \textsc{Matsumoto}\altaffilmark{1,2}, Shuji \textsc{Matsuura}\altaffilmark{1}, Itsuki \textsc{Sakon}\altaffilmark{3}, Masahiro \textsc{Tanaka}\altaffilmark{4}, and  Takehiko \textsc{Wada}\altaffilmark{1}}
\altaffiltext{1}{Department of Space Astronomy and Astrophysics, Institute of Space and Astronautical Science, Japan Aerospace Exploration Agency, 3-1-1 Yoshinodai, Chuo-ku, Sagamihara, Kanagawa 252-5210}
\altaffiltext{2}{Institute of Astronomy and Astrophysics, Academia Sinica, No.1, Roosevelt Rd, Sec. 4, Taipei 10617, Taiwan, R.O.C.}
\altaffiltext{3}{Department of Astronomy, Graduate School of Science, The University of Tokyo, Hongo 7-3-1, Bunkyo-ku, Tokyo 113-0033}
\altaffiltext{4}{Center for Computational Sciences, University of Tsukuba, 1-1-1 Tennodai, Tsukuba, Ibaraki 305-8577}
\email{tsumura@ir.isas.jaxa.jp}
\maketitle
 
\begin{abstract}
We first obtained the spectrum of the diffuse Galactic light (DGL) at general interstellar space in 1.8-5.3 $\mu$m wavelength region
with the low-resolution prism spectroscopy mode of the AKARI Infra-Red Camera (IRC) NIR channel.
The 3.3 $\mu$m PAH band is detected in the DGL spectrum at Galactic latitude $\mid b \mid < 15^{\circ }$, and its correlations with the Galactic dust and gas are confirmed.
The correlation between the 3.3 $\mu$m PAH band and the thermal emission from the Galactic dust is expressed not by a simple linear correlation but by a relation with extinction.
Using this correlation, the spectral shape of DGL at optically thin region ($5^{\circ } < \mid b \mid < 15^{\circ }$) was derived as a template spectrum.
Assuming that the spectral shape of this template spectrum is uniform at any position, 
DGL spectrum can be estimated by scaling this template spectrum using the correlation between the 3.3 $\mu$m PAH band and the thermal emission from the Galactic dust.
\end{abstract}

\section{Introduction}
The Diffuse Galactic Light (DGL) comprises scattered starlight by dust particles in the interstellar space at $<$3 $\mu$m, 
and emissions from the dust particles with some band features at longer wavelengths\footnote{Although the term DGL sometimes indicates only the scattered starlight component, the term DGL indicates both scattered starlight and emission components in this paper.}.
Thus observational studies of DGL is important to investigate the dust property in our Galaxy, 
and it is also important for deriving the extragalactic background light (EBL) since DGL is one of foregrounds for the EBL measurement.
However, isolation of DGL from other diffuse emissions, especially the strongest zodiacal light (ZL) foreground, is very difficult due to its diffuse, extended nature.
A commonly-used method to estimate DGL is the correlation with the dust column density estimated by the thermal emission of the dust from far-infrared (100 $\mu$m) observations, 
or the column density of HI and/or CO from radio observations.
In the optical wavelengths, DGL brightness \citep{Witt2008, Matsuoka2011, Ienaka2013} and spectrum \citep{Brandt2012}  are obtained by the correlation with the 100 $\mu$m dust thermal emission.
However, observations of DGL at near-infrared (NIR) are limited and controversial.

The presence of the infrared band features in DGL has been first confirmed for the 3.3 $\mu$m band by the AROME balloon experiment \citep{Giard1988}.
Such ubiquitous Unidentified Infrared (UIR) bands are a series of distinct emission bands seen at 3.3, 5.3, 6.2, 7.7, 8.6, 11.2, and 12.7 $\mu$m,
and they are supposed to be carried by the polycyclic aromatic hydrocarbons (PAH) \citep{Leger1984, Allamandola1985}.
They are excited by absorbing a single ultraviolet (UV) photon and release the energy with a number of infrared photons in cascade 
via several lattice vibration modes of aromatic C-H and C-C bonds \citep{Allamandola1989}.
The 3.3 $\mu$m PAH band emission has been assigned to the stretching mode transition ($v=1-0$) of the C-H bond on aromatic rings.
There is a quantitative model for DGL from interstellar dust including PAH \citep{Li2001}.
The correlation between the 3.3 $\mu$m PAH band detected by the Near-Infrared Spectrometer (NIRS) on the Infrared Telescope in Space (IRTS) 
and  the 100 $\mu$m thermal emission of the large dust grains by the Infrared Astronomical Satellite (IRAS) was confirmed 
at the Galactic plane region ($42^{\circ } < l < 55^{\circ }$, $\mid b \mid < 5^{\circ }$), 
implying that the PAH molecules are well mixed with the large dust grains at the Galactic plane \citep{Tanaka1996}.

\begin{figure*}
  \begin{center}
    \FigureFile(160mm,100mm){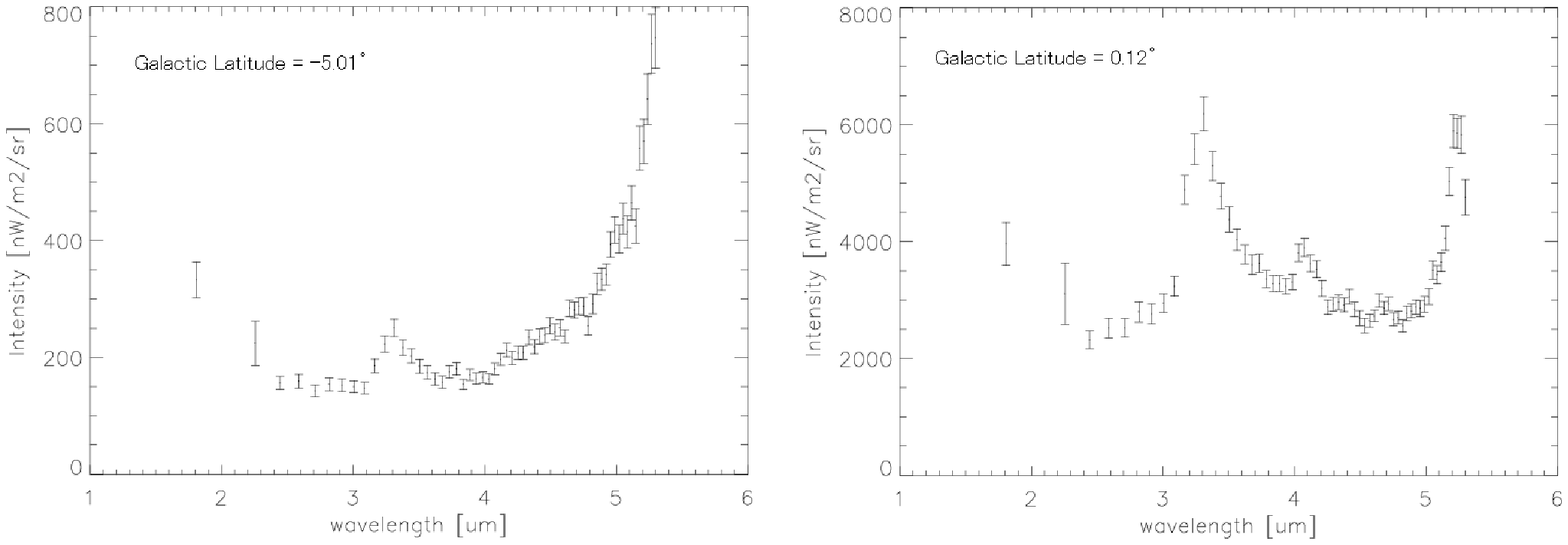}
  \end{center}
  \caption{Examples of the diffuse sky spectra obtained with AKARI IRC. (a) Spectrum at $b = -5^{\circ }$. The 3.3 $\mu$m PAH band is clearly visible on the bottom of the ZL spectrum.
           (b) Spectrum at the Galactic plane. DGL dominates the sky brightness and strong PAH bands at 3.3 $\mu$m and 5.25 $\mu$m and Br-$\alpha $ at 4.1 $\mu$m are clearly visible.}
  \label{spectrum}
\end{figure*}

In this paper, we describe DGL spectrum obtained with the low-resolution prism spectroscopy mode on the AKARI Infra-Red Camera (IRC) NIR channel in 1.8-5.3 $\mu$m wavelength region.
Our idea to derive DGL spectrum at NIR in this paper is to use the 3.3 $\mu$m PAH band feature as a tracer of DGL 
combined with the correlation with the 100 $\mu$m dust thermal emission.
The 3.3 $\mu$m PAH band is detected in this wavelength region at $\mid b \mid < 15^{\circ }$, 
and the correlation with the thermal emission of the large dust grains is also confirmed.
Using this correlation, we developed a method to estimate the DGL spectrum at NIR at any location. 

This paper is organized as follows. 
In Section \ref{sec_reduction}, we describe the data reduction. 
In Section \ref{sec_PAH}, we describe the correlation of the 3.3 $\mu$m PAH band feature in DGL with Galactic latitude and the distribution of Galactic dust and gas.
The method to estimate DGL spectrum using this correlation is shown in Section \ref{sec_DGL}, and the summary of this paper is given in Section \ref{sec_summary}.
There are two companion papers describing the spectrum of the infrared diffuse sky; 
ZL is described in \citet{Tsumura2013a} (hereafter Paper I) and EBL is described in \citet{Tsumura2013c} (hereafter Paper III) 
in which the foregrounds described in Paper I and this paper (Paper II) are subtracted.

\section{Data Selection and Reduction} \label{sec_reduction}
AKARI is the first Japanese infrared astronomical satellite launched at February 2006, equipped with a cryogenically cooled telescope of 68.5 cm aperture diameter \citep{Murakami07}.
IRC is one of two astronomical instruments of AKARI, and it covers 1.8-5.3 $\mu$m wavelength region with a 512$\times $412 InSb detector array in the NIR channel\footnote{IRC has two other channels covering 5.8-14.1 $\mu$m in the MIR-S channel and 12.4-26.5 $\mu$m in the MIR-L channel.} \citep{Onaka07}.
It provides low-resolution ($\lambda /\Delta \lambda \sim 20$) slit spectroscopy for the diffuse radiation by a prism\footnote{High-resolution spectroscopy ($\lambda /\Delta \lambda \sim 120$) with a grism is also available.} \citep{Ohyama2007}.
One biggest advantage to the previous IRTS measurement \citep{Tanaka1996} is the higher angler resolution (1.46 arcseconds) of AKARI IRC which allows us to detect and remove fainter point sources,
while the IRTS measurement was highly contaminated by bright stars at the Galactic plane because of its resolution (8 arcminutes).

See Paper I for the details of the data selection and reduction.
Here we simply note that 278 pointed data of diffuse spectrum were selected in this study distributing wide range of ecliptic and Galactic coordinates.
Dark current was subtracted by the method specialized for the diffuse sky analysis described in \citet{TsumuraWada2011}. 
Stars brighter than $m_K(\textrm{Vega}) = 19$ were detected on the slit and masked for deriving the diffuse spectrum.
It was confirmed that the brightness due to unresolved Galactic stars under this detection limit is negligible ($<$0.5 \% of the sky brightness at 2.2 $\mu$m) 
by a Milky Way star counts model, TRILEGAL \citep{Girardi2005}.
Cumulative brightness contributed by unresolved galaxies can be estimated by the deep galaxy counts, being $<$4 \nw\ at K band in the case of limiting magnitude of $m_K = 19$ \citep{Keenan10}.

\section{3.3 $\mu$m PAH band} \label{sec_PAH}
\subsection{Association to our Galaxy}
Figure \ref{spectrum} shows examples of the spectra of the infrared diffuse sky used in this study.
Although the obtained spectra are dominated by ZL except for the Galactic plane, the 3.3 $\mu$m PAH band is detectable at $\mid b \mid < 15^{\circ }$ in our dataset.
The 3.3 $\mu$m PAH band is easy to be found at the bottom of the ZL spectrum because ZL spectrum has a local minimum at around 3.5 $\mu$m as shown in Figure \ref{spectrum} (a).
At the Galactic plane, DGL dominates the sky spectrum as shown in Figure \ref{spectrum} (b).
The spectral shape of the 3.3 $\mu$m band is asymmetry because other 3.4 and 3.5 $\mu$m PAH sub-band features are combined and detected together,
which were separately detected by the high-resolution spectroscopy ($\lambda /\Delta \lambda \sim 120$) with the IRC grism mode at the Galactic plane \citep{Onaka2011}.

The 3.3 $\mu$m PAH band was extracted from the sky spectrum by almost the same method used in \citet{Tanaka1996}.
First, the continuum intensity at 3.3 $\mu$m ($\lambda I_{3.3\mu m}^{cont}$) was interpolated between 3.2 and 3.8 $\mu$m.
Although the continuum was interpolated between 3.2 and 3.6 $\mu$m in \citet{Tanaka1996}, 
we used the intensity at 3.8 $\mu$m for the interpolation to avoid the contamination from the PAH sub-feature at 3.5 $\mu$m.
Then, the total energy of the 3.3 $\mu$m PAH band feature ($E_{3.3}$) was calculated as the excess from the continuum;
\begin{equation} E_{3.3} = \frac{\Delta \lambda }{\lambda } [\lambda I_{3.3\mu m} - \lambda I_{3.3\mu m}^{cont})]+0.58  \end{equation}
where $\Delta \lambda$ = 0.13 $\mu$m was employed for direct comparison to \citet{Tanaka1996}, 
and two data points around 3.3 $\mu$m were summed to compute $\lambda I_{3.3\mu m}$ for matching the wavelength resolution to IRTS and reducing the error. 
A small offset of 0.58 \nw\ is applied to correct the difference between the ZL continuum and the linear interpolation between 3.2 and 3.8 $\mu$m.
PAH band at $\mid b \mid < 15^{\circ }$ was detected by this method.

\begin{figure}
  \begin{center}
    \FigureFile(80mm,50mm){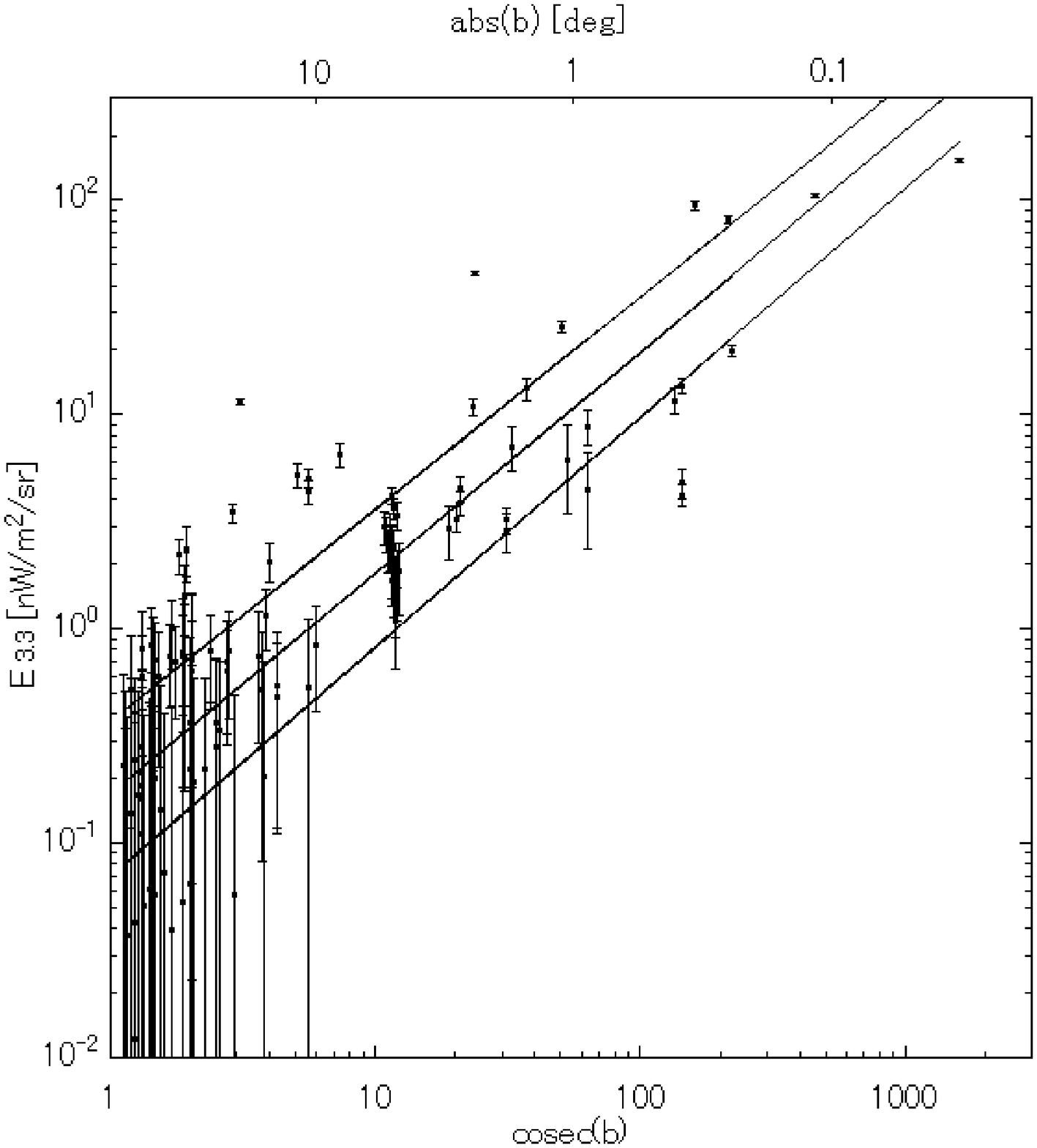}
  \end{center}
  \caption{The correlation of the 3.3 $\mu$m PAH band intensity ($E_{3.3}$) to the Galactic latitude ($b$). The solid lines are the best-fit lines to the data.}
  \label{cosec}
\end{figure}

First, we show a general correlation of this PAH band with our Galaxy in Figure \ref{cosec}. 
It means that the PAH dust is associated with DGL from our Galaxy, and this correlation can be expressed by
\begin{equation} E_{3.3} = (0.17^{+0.2}_{-0.1}) \cdot (\textrm{cosec} \mid b \mid ) ^{(1.03\mp 0.04)} \end{equation}
Since such correlation between the 3.3 $\mu$m PAH band and ecliptic latitude was not detected, 
we can conclude that the observed PAH features is not associated with ZL from the Solar system.

\subsection{Association to the distribution of dust and gas}
Next, we tested the correlations of the PAH  band with the distribution of Galactic dust and gas, which are not correlated simply with Galactic latitude. 
The 100 $\mu$m intensity map ($\lambda I_{100 \mu m}$) that is a reprocessed composite of the COBE/DIRBE and IRAS/ISSA maps (SFD map, see \citet{Schlegel1998}) 
is used as the dust distribution,
and the column density map of HI obtained from the Leiden/Argentine/Bonn (LAB) Galactic HI survey \citep{Kalberla2005} is used as the gas distribution map.
The good correlation between the dust and gas was reported \citep{Stark92, Arendt98},
and HI column density can be converted to the 100 $\mu$m intensity by $0.018 \pm 0.003$ (\uw)/($10^{20}$ atoms/cm$^2$) at HI $< 10^{22}$ atoms/cm$^2$ as adopted in \citet{Matsuura2011}. 
In previous works, the correlation between the 3.3 $\mu$m PAH band and the 100 $\mu$m thermal intensity was reported by \citet{Giard1989} and \citet{Tanaka1996} 
but they were limited at the Galactic plane ($\mid b \mid < 6^{\circ }$).
AKARI data in this study provides the correlations up to $\mid b \mid = 15^{\circ }$ with higher angler resolution and higher sensitivity of point sources to remove foreground stars than the previous works.

\begin{figure*}
  \begin{center}
    \FigureFile(160mm,100mm){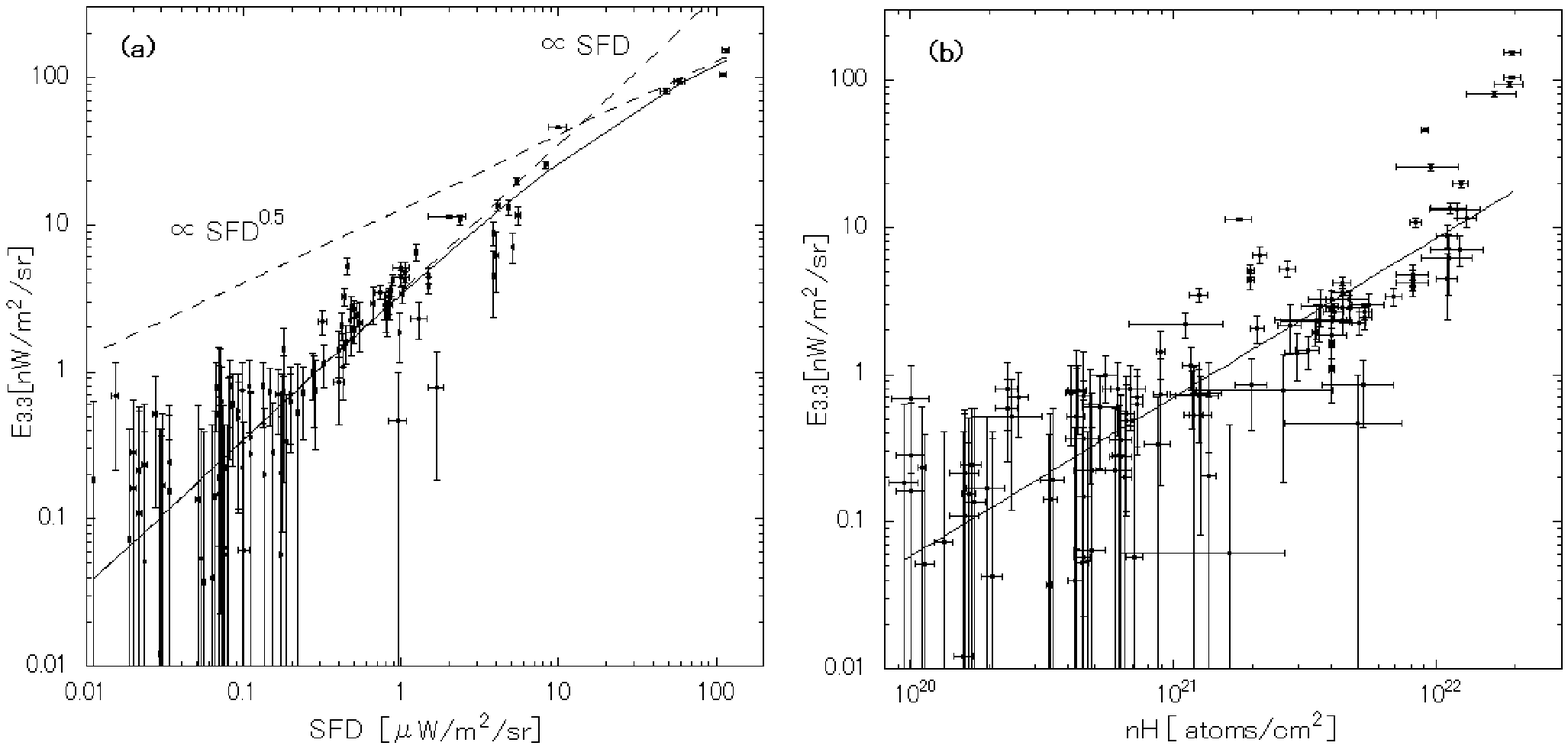}
  \end{center}
  \caption{The correlation of the PAH 3.3 $\mu$m band intensity with (a) the 100 $\mu$m intensity from the SFD map \citep{Schlegel1998}  and (b) column density of HI from the LAB survey \citep{Kalberla2005}.
           Best fit curves of equation (\ref{eq_SFD}) for figure (a) and equation (\ref{eq_nH}) for figure (b) are also shown by solid curves.
           Two extreme cases of $\propto SFD$ and $\propto \sqrt{SFD}$ are also shown by broken lines in figure (a).}
  \label{E33}
\end{figure*}

Figure \ref{E33} shows the correlations of the 3.3 $\mu$m PAH band of our data set with the 100 $\mu$m thermal intensity from the SFD map \citep{Schlegel1998}
and the column density of HI from the LAB survey \citep{Kalberla2005}.
These correlations can be expressed by
\begin{equation} \frac{E_{3.3}}{\textrm{\nw}} = (2.9 \pm 1.0) \cdot  \left( \frac{\lambda I_{100 \mu m}}{\textrm{\uw}}\right)^{(0.91\mp 0.04)}  \label{eq_correlation}\end{equation}
\begin{equation} \frac{E_{3.3}}{\textrm{\nw}} = (0.7 \pm 0.3) \cdot  \left( \frac{\textrm{HI}}{10^{21} \ \textrm{atoms/cm}^2}\right)^{(1.08\mp 0.05)}  \label{eq_nH} \end{equation}
These correlations are better than the correlation with the Galactic latitude shown in Figure \ref{cosec},
and it means that PAH molecules, interstellar dust, and interstellar gas are well mixed in the interstellar space.
Deviation of the correlation with HI in LAB survey data shown in Figure \ref{E33} (b)  is larger than the correlation with the 100 $\mu$m intensity in SFD map shown in Figure \ref{E33} (a),
because angular resolution of SFD map is better than LAB map allowing us better point-to-point correlation analysis. 
Thus advanced analysis described in the next sub-section is investigated only for the correlation with the 100 $\mu$m intensity.

\subsection{The effect of extinction on the PAH band}
\citet{Tanaka1996} assumed a simple linear relation between $E_{3.3}$ and $\lambda I_{100 \mu m}$ 
obtaining a relation of $E_{3.3}/\lambda I_{100 \mu m}= (2.9\pm 0.9)\times 10^{-3}$, 
and systematic difference from the linear relation at high $\lambda I_{100 \mu m}$ region was concluded to be due to extinction.  
\citet{Giard1989} also investigated the correlation between $E_{3.3}$ and $\lambda I_{100 \mu m}$ and found the extinction at the Galactic plane.
The power of $0.91 \mp 0.04$ (smaller than unity) in our fitting in equation (\ref{eq_correlation}) is consistent with the extinction as mentioned in \citet{Giard1989} and \citet{Tanaka1996}.
Therefore, we conducted a fitting including the effect of extinction.

When the source's term in the transfer equation is proportional to extinction term all along the line of sight,
extinction can be written as $(1-e^{-\tau_{\lambda}})/\tau_{\lambda}$, where $\tau_{\lambda}$ is the optical depth at a wavelength ${\lambda}$ \citep{Giard1989}.
Then the correlation with extinction between the PAH emission ($E_{3.3}$) and the  the 100 $\mu$m intensity ($\lambda I_{100 \mu m}$) can be written as 
\begin{equation}   \frac{E_{3.3}}{\textrm{\nw}} = \alpha \frac{1-e^{-\tau_{3.3} }}{\tau_{3.3} } \left( \frac{\lambda I_{100 \mu m}}{\textrm{\uw}} \right ) \label{eq_SFD} \end{equation}
where $\alpha$ is a fitting parameter.
Assuming that the 100 $\mu$m intensity is proportional to the number of PAH molecules associating with dust particles, and the number of dust particles is proportional to $\tau + \tau^2$ \citep{Rybicki1986},
thus the optical depth $\tau_{3.3}$ can be obtained as a solution of this equation.
\begin{equation}  \frac{\lambda I_{100 \mu m}}{\textrm{\uw}} = \frac{1}{\beta}  (\tau_{3.3} +\tau_{3.3} ^2)  \label{eq_tau}  \end{equation}
where $\beta$ is another fitting parameter.
By fitting to our data, we obtained $\alpha =3.5 \pm 1.0$ and $\beta =0.10 \pm 0.04$.
The curve of equation (\ref{eq_SFD}) with best fit parameters is shown as a solid curve in Figure \ref{E33} (a).

In the optically thin case ($\tau_{3.3} \ll 1$), the extinction term $(1-e^{-\tau_{3.3}})/\tau_{3.3}$ becomes unity.
Thus we obtain the linear correlation.
\begin{equation}  \frac{E_{3.3}}{\textrm{\nw}} = \alpha \left( \frac{\lambda I_{100 \mu m}}{\textrm{\uw}} \right) \; \; \;  (\tau_{3.3} \ll 1) \end{equation}
In the optically thick case ($\tau_{3.3} \gg 1$), the extinction term can be written as $1/\tau_{3.3}$ because the term $e^{-\tau_{3.3}}$ becomes zero.
In addition, the optical depth $\tau_{3.3}$ can be written $\tau_{3.3} = \sqrt{\beta \cdot \lambda I_{100 \mu m}}$ from equation (\ref{eq_tau}) owing to $\tau_{3.3} \gg 1$.
Combining these equations, we obtain
\begin{equation}  \frac{E_{3.3}}{\textrm{\nw}} = \frac{\alpha}{\sqrt{\beta}} \sqrt{ \frac{\lambda I_{100 \mu m}}{\textrm{\uw}} } \; \; \;  (\tau_{3.3} \gg 1)  \end{equation}
These two extreme cases are also shown as broken lines in Figure \ref{E33} (a).

The gradient of $\alpha =3.5 \pm 1.0$ is higher than the value from the IRAS result of  $\alpha = 2.5 \pm 0.4$ \citep{Giard1994},
but this result was determined based on the data averaged in the Galactic latitude range of $\mid b \mid  < 1^{\circ }$ where the bright discrete sources are included.
IRTS, with higher sensitivity than IRAS but still limited at  $\mid b \mid  < 5^{\circ }$, obtained the value of $\alpha = 2.9 \pm 0.9$ in \citet{Tanaka1996}, which is closer to our result. 
The 3.3 $\mu$m PAH band intensity deviates from the linearity at  $\lambda I_{100 \mu m} > 10$ \uw\ or $\mid b \mid  < 1^{\circ }$.
This is equivalent to $\tau _{3.3}=0.6$ at  $\lambda I_{100 \mu m} > 10$ \uw\ or $\mid b \mid  < 1^{\circ }$ in our fitting, 
which is higher than the estimated value of $\tau _{3.3}=0.18$ at $\mid b \mid  < 0.75^{\circ }$ based on the extinction low summarized in \citet{Mathis1990} and optical depth at 240 $\mu$m from \citet{Sodroski1994}.

\begin{figure*}
  \begin{center}
    \FigureFile(160mm,100mm){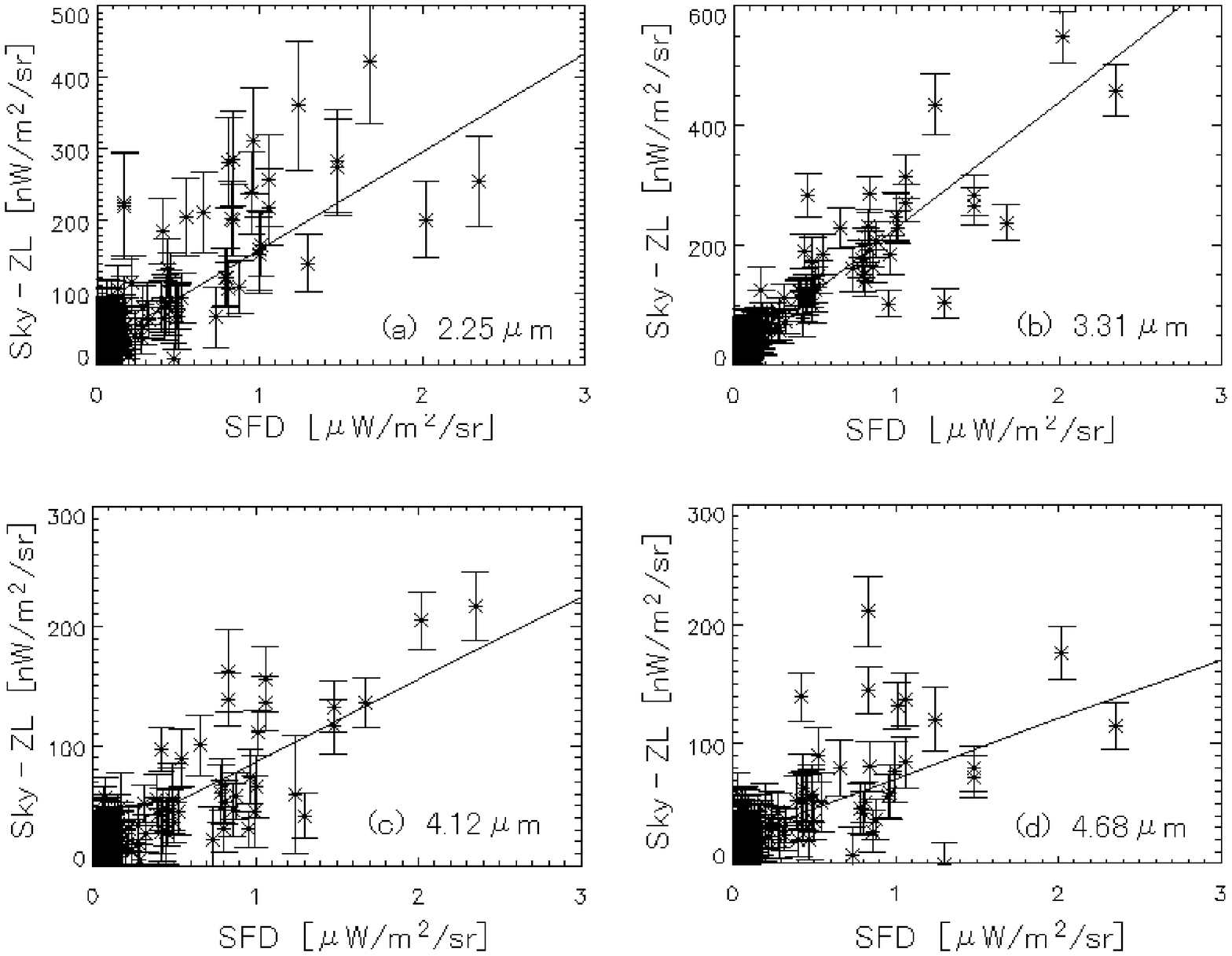}
  \end{center}
  \caption{Examples of correlation between $SKY(\lambda ) - ZL(\lambda )$ and the 100 $\mu$m intensity in SFD map at  $ \lambda I_{100 \mu m} < 3$ \nw\  at (a) 2.25 $\mu$m, (b) 3.31 $\mu$m, (c) 4.12 $\mu$m, and (d) 4.68 $\mu$m. }
  \label{correlation}
\end{figure*}

\section{DGL spectrum} \label{sec_DGL}
\subsection{Correlation method}
\begin{figure*}
  \begin{center}
    \FigureFile(160mm,100mm){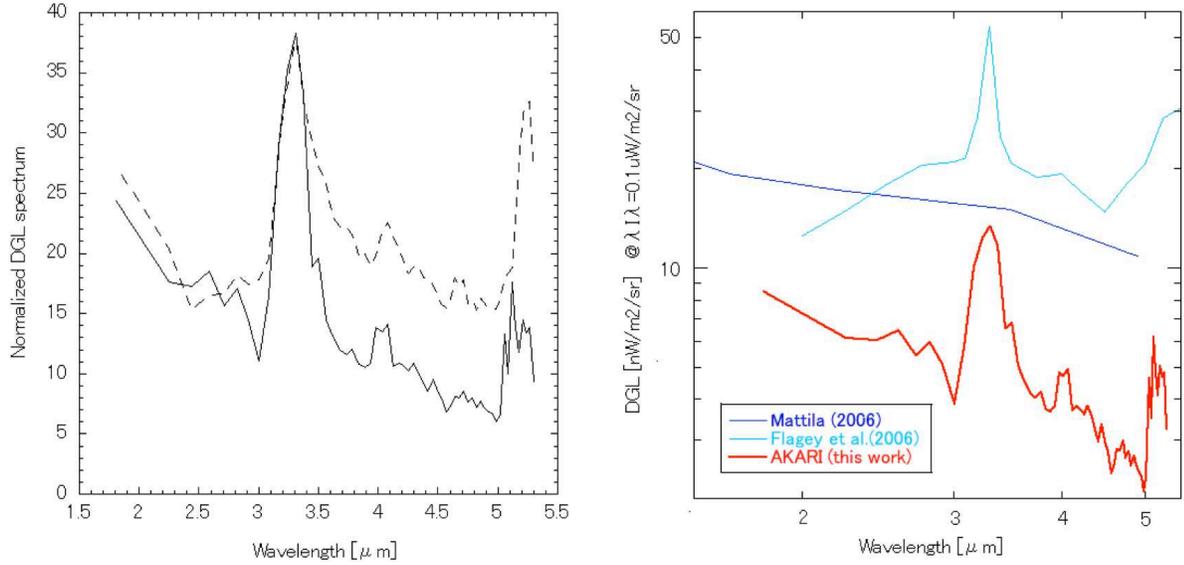}
  \end{center}
  \caption{(a) DGL template spectrum (solid curve) from the correlation of the data at $5^{\circ }< \mid b \mid <15^{\circ }$
  and the averaged spectrum at Galactic plane (broken curve) normalized to be $E_{3.3}=1$. 
  The error of the obtained DGL spectrum by this correlation method is 5\% at $<3.8 \mu$m, 15\% between $3.8 \mu$m and $4.2 \mu$m, and 20\% at $>4.2 \mu$m.           
  (b) DGL spectra scaling to $\lambda I_{100 \mu m}= 0.1$ \uw\ or HI = $5 \times 10^{20}$ atoms/cm$^2$ by our method and other estimations \citep{Mattila2006, Flagey2006}.}
  \label{DGL}
\end{figure*}

In this section, we develop a method to derive the DGL spectrum at 1.8-5.3 $\mu$m by the correlation with the 100 $\mu$m intensity.
The diffuse sky spectrum includes ZL, DGL, and EBL, i.e.
\begin{equation} SKY (\lambda ) = ZL (\lambda ) + DGL (\lambda ) +EBL(\lambda )  \end{equation}
ZL is modeled in Paper I which can be subtracted based on the DIRBE ZL model \citep{Kelsall98}, and EBL is the isotropic component.
Therefore, only DGL has the correlation with the 100 $\mu$m intensity,
so DGL can be derived by the correlation by assuming a linear correlation with the 100 $\mu$m intensity.
\begin{equation} SKY(\lambda ) - ZL(\lambda ) = a(\lambda ) \cdot \lambda I_{100 \mu m} + b(\lambda ) \end{equation}
where $ a(\lambda ) \cdot \lambda I_{100 \mu m} $ is equivalent to $DGL(\lambda )$ and $b(\lambda )$ is equivalent to $EBL(\lambda )$.
Figure \ref{correlation} shows the correlation between $SKY(\lambda ) - ZL(\lambda )$ in our dataset and the 100 $\mu$m intensity from the SFD map \citep{Schlegel1998}.
The data for this correlation analysis were selected by a criteria of $ \lambda I_{100 \mu m} < 3$ \uw, equivalent to the Galactic latitude $\mid b \mid >5^{\circ }$,  
to trace low dust density regions owing to the assumption of the linear correlation.
The gradients as a function of wavelength in Figure \ref{correlation}, $a(\lambda )$, correspond to the spectral shape of DGL.
Normalized spectrum of the obtained DGL spectrum is shown in Figure \ref{DGL} (a), and the 3.3 $\mu$m PAH band feature in DGL was detected.
The error of the obtained DGL spectrum by this correlation method is 5 \% at $<$3.8 $\mu$m, 15 \% between 3.8 $\mu$m and 4.2 $\mu$m, and 20 \% at $>$4.2 $\mu$m.
Since the spectral shape of DGL may vary depending on environments, it is a representative spectrum of DGL at low dust density regions, typically $5^{\circ }< \mid b \mid <15^{\circ }$,
and the variance of environments is included in these errors.

In this correlation method, we assumed the linear correlation between DGL and the 100 $\mu$m intensity,
but we have already showed the non-linear correlation between the 3.3 $\mu$m PAH band and the 100 $\mu$m intensity owing to extinction as shown in Figure \ref{E33} (a).
Thus we modify this DGL estimation method by combining it with the 3.3 $\mu$m PAH band as a tracer for scaling the DGL spectrum at any general sky.
First, the template DGL spectrum $DGL_{\textrm{temp}}(\lambda )$ is defined as the derived DGL spectrum by this correlation method normalized to be $E_{3.3}=1$,
\begin{equation} DGL_{\textrm{temp}}(\lambda) = \frac{a(\lambda)}{E_{3.3}(a(\lambda))}    \end{equation}
This template DGL spectrum is shown in Figure \ref{DGL} (a).
Assuming that the spectral shape of this template DGL spectrum does not change at any location, 
we can estimate the DGL spectrum at any place by scaling this template DGL spectrum by $E_{3.3}$ 
which can be obtained as a function of the 100 $\mu$m thermal intensity using the correlation shown in Figure \ref{E33} (a), i.e.,
\begin{equation}  DGL(\lambda ) =  E_{3.3}(I_{100 \mu m}) \cdot DGL_{\textrm{temp}}(\lambda )  \end{equation}
Figure \ref{DGL} (b) shows the the resultant DGL spectrum with other DGL estimations \citep{Mattila2006, Flagey2006} scaling to $\lambda I_{100 \mu m}= 0.1$ \uw\ or HI = $5 \times 10^{20}$ atoms/cm$^2$.

\begin{table*}
\begin{center}
\caption{Selected regions at the Galactic plane for evaluating the DGL spectrum}
\label{DGL_table}
 \begin{tabular}{ccccc}
   \hline
   Pointing ID & Galactic longitude $l$ & Galactic latitude $b$ & $\lambda I_{100 \mu m}$  \uw & HI atoms/cm$^2$ \\
   \hline
   410017.1 & 30.76 & 0.36 & 59.00 & $1.89 \times 10^{22}$ \\
   410018.1 & 31.01 & 0.12 & 111.09 & $1.95 \times 10^{22}$ \\
   410021.1 & 30.99 & -0.04 & 115.25 & $1.95 \times 10^{22}$ \\
   410022.1 & 31.24 & -0.27 & 47.65 & $1.66 \times 10^{22}$\\
   \hline
\end{tabular}
\end{center}
\end{table*}

Our DGL spectrum is lower than the other previous estimations, but our method gives a better DGL estimation at the general interstellar space with low dust density.
In the previous works, DGL was estimated by scaling using HI based on only a limited number of specific dense regions with HI $\sim 2 \times 10^{22}$ atoms/cm$^2$.
However, the ratio of DGL/HI at dense regions ($>10^{22}$ atoms/cm$^2$) is higher than that in the general interstellar fields as shown in Figure \ref{E33} (b), 
which leads high DGL estimations in the previous works.
On the other hand, our estimation is based on a number of the wide-spread data points at the general interstellar fields with low dust density with higher spacial resolution to remove stars, and scaling is based on the 100 $\mu$m intensity which has tighter correlation to DGL as shown in Figure \ref{E33} (a).
Therefore our estimate gives more reliable estimation of the DGL spectrum especially at low dust density regions.

Although our DGL estimation is lower than previous estimations, 
it may still overestimates the 3.3 $\mu$m intensity at high Galactic latitude regions.
The 3.3 $\mu$m PAH band was detected only at the region of  $\mid b \mid <15^{\circ }$ in our dataset, 
and we assumed that the obtained spectral shape of DGL does not change at any location in this method, but there is no guarantee that this assumption is valid at high Galactic latitude regions.
For example, UV radiation field at high Galactic latitude regions is weaker than that at Galactic plane \citep{Seon2011}, 
therefore the PAH molecules are less excited at high Galactic latitude than Galactic plane. 
In such a case, our method overestimates the 3.3 $\mu$m intensity in DGL at high Galactic latitude as implied in Paper III.

\subsection{DGL spectrum in the Galactic plane}
We compared the obtained DGL spectrum by the correlation method ($5^{\circ } < \mid b \mid < 15^{\circ }$) with the spectrum at the Galactic plane, 
because the diffuse sky spectrum at the Galactic plane is dominated by DGL. 
For example, DGL brightness at Galactic plane in this wavelength region is several thousands \nw, while ZL brightness is a few hundreds \nw, which is less than 5 \% as shown in Figure \ref{spectrum}.
Therefore, the spectral shape of DGL can be evaluated by the diffuse sky spectra at the Galactic plane.
We selected four spectra with the strongest 100 $\mu$m dust thermal emission and HI column density in our data set, located at the Galactic plane ($l$, $b$)$\sim$(31$^{\circ }$, 0$^{\circ }$) summarized in Table \ref{DGL_table}, where the brightness of ZL contribution is $<$5 \%. 
These spectra at the Galactic plane are similar with each other with $\sim$15 \% dispersion, 
and the broken curve in Figure \ref{DGL} (a) shows the average spectrum of  these selected spectra normalized to be $E_{3.3}=1$.
The 3.3 $\mu$m PAH band is the most distinctive among others, and the second outstanding feature at 5.25 $\mu$m is also the PAH feature \citep{Allamandola1989b, Cohen1989, Boersma2009}.
Br-$\alpha $ at 4.05 $\mu$m and Pf-$\beta $ at 4.65 $\mu$m are also detected, which can be useful information to estimate ionizing temperature and extinction.
One example of such study in M17 case with AKARI IRC high-resolution spectroscopy mode ($\lambda /\Delta \lambda \sim 120$) can be found in \citet{Onaka2011}.

An excess continuum emission at the Galactic plane was confirmed at $>$3.5 $\mu$m as shown in Figure \ref{DGL} (a).
This excess continuum emission was first reported in visual reflection nebulae \citep{Sellgren1983}, and then found in galaxies \citep{Lu2003, Onaka2010} and DGL \citep{Flagey2006}.
The emission process of this excess continuum is still unknown, but \citet{Flagey2006} suggests the PAH fluorescence excited by UV photons.
This excess continuum emission may be a reason why the previous works overestimated DGL from high dust density regions as shown in Figure \ref{DGL} (b).

\section{Summary} \label{sec_summary}
The 3.3 $\mu$m PAH band is detected in the diffuse sky spectrum of the interstellar space at $\mid b \mid < 15^{\circ }$, 
and this band intensity is correlated with the 100 $\mu$m thermal intensity of interstellar dust and HI column density.
We modeled the correlation between the 3.3 $\mu$m PAH band and the 100 $\mu$m thermal intensity with extinction.
We also introduce a method to estimate the DGL spectrum at 1.8-5.3 $\mu$m.
This is the first estimation of DGL spectrum in the general sky at NIR based on observation.
In this method, the spectral shape of DGL is derived by the correlation with the 100 $\mu$m thermal intensity,
and it is scaled by the correlation of the 3.3 $\mu$m PAH band brightness as a tracer.
DGL spectrum estimated by our method is lower than the previous estimations, but our result is more reliable for the regions with low dust density regions
because our result is based on a wide range of general interstellar field although previous result is based on some specific region with high dust density.
In addition, we found the excess continuum emission at Galactic plane at 3-5 $\mu$m as reported by previous works.

\bigskip
This research is based on observations with AKARI, a JAXA project with the participation of ESA.
This research is also based on significant contributions of the IRC team.
We thank Dr. Mori-Ito Tamami (The university of Tokyo) and Mr. Arimatsu Ko (ISAS/JAXA) for discussion about PAH.
The authors acknowledge support from Japan Society for the Promotion of Science, KAKENHI (grant number 21111004 and 24111717).

\end{document}